\documentclass[12pt]{article}
\usepackage{jheppub}
\usepackage{amssymb,amsmath, graphicx}
\usepackage{latexsym}
\usepackage{mathrsfs}
\usepackage{hyperref} 
\usepackage{verbatim}
\usepackage{bm}
\usepackage{mathtools}

\usepackage{xcolor}

\def\bea{\begin{eqnarray}}
\def\eea{\end{eqnarray}}
\def\be{\begin{equation}}
\def\ee{\end{equation}}
\def\ba{\begin{array}}
\def\ea{\end{array}}

\newcommand{\calR}{{\cal R}}

\newcommand{\calL}{{\cal L}}
\newcommand{\calH}{{\cal H}}

\newcommand{\Mpl}{{M_{\mathrm{Pl}}^2}}

\newcommand{\mpl}{{M_{\mathrm{Pl}}}}
\newcommand{\de}{{\mathrm{d}}}

\def\d{{\rm d}}

\def\C{{\cal C}}
\def\0{{\boldsymbol 0}}
\def\e{{\boldsymbol e}}
\def\k{{\boldsymbol{k}}}
\def\m{{\boldsymbol{m}}}

\def\v{{\boldsymbol{v}}}
\def\x{{\vec{x}}}

\def\e{{\boldsymbol{e}}}

\begin{document}
\title{Disformal transformations on the CMB}
\author[a]{Clare Burrage,}
\affiliation[a]{
School of Physics and Astronomy, University of Nottingham, Nottingham, NG7 2RD, UK}
\author[b]{Sebastian Cespedes}
\author[b]{and\ Anne-Christine Davis}
\affiliation[b]{DAMTP, University of Cambridge, Wilberforce Road, Cambridge, CB3 0WA, UK}
\emailAdd{scespedes@damtp.cam.ac.uk}
\emailAdd{clare.burrage@nottingham.ac.uk}
\emailAdd{a.c.davis@damtp.cam.ac.uk}
\abstract{In this work we study the role of disformal transformation on  cosmological backgrounds and its relation to the speed of sound for tensor modes.  A speed different from one for tensor modes can arise in several contexts, such as Galileons theories or massive gravity, nevertheless the speed is very constrained to be one by observations of gravitational wave emission.  It has been shown that in inflation a disformal   transformation allows to set the speed for  tensor modes to  one without making  changes to the curvature power spectrum. Here we show that this invariance does not hold when considering the CMB anisotropy power spectrum. It turns out that the after doing the transformation there is an imprint on the acoustic peaks and  the diffusion damping. This has interesting consequences; here we explore quartic galileon theories which allow a modified speed for tensor modes.  For these theories the transformation can be used to constraint  the parameter space in different regimes. } 
\maketitle
\newpage
\section{Introduction}
It is compelling to use the knowledge of gravitational waves to try to learn more about primordial cosmology. With the first detection of LIGO and the future experiments on primordial gravitational waves to come~\cite{Abbott:2016blz,Accadia:2011zzc, Somiya:2011np,  Kawamura:2011zz, AmaroSeoane:2012km}, it is interesting to investigate what we can learn about primordial cosmology from them. One of the advantages of using gravitational waves is that it allows us to make bold model independent predictions for cosmology.  For example the speed  of tensor modes  $c_T$ has been  strongly constrained to be  $(c_T-1)<10^{-7}$~\cite{Nishizawa:2014zna, Calabrese:2016bnu, Blas:2016qmn}. This fact can be used to study thoroughly the validity of general relativity at early times. 

It is well known that  general relativity has passed all experimental tests with flying colours~\cite{Koyama:2015vza}. Nevertheless there are many contexts in which it might be necessary to consider its possible modifications~\cite{Clifton:2011jh}. For example, during inflation higher curvature corrections could be the responsible producing the primordial fluctuations power spectrum. This could be seen as higher order operators overtaking the Einstein Hilbert action  hence modifying  the equations of motion at tree level~\cite{Cheung:2007st}. Furthermore, the prevalence of dark energy still remains unexplained and thus whether it could be solved by modifying general relativity at late times remains a possibility~\cite{Copeland:2006wr}.  Although most of these theories are very different in nature, they share similar predictions. In particular a modification of the speed of propagation of tensor modes can arise quite generally: for example when there is more than one metric evolving~\cite{Magueijo:2003gj} or also when higher order operators  are enhanced by the requirement of certain symmetries~\cite{Cheung:2007st, Nicolis:2008in}. This leads to a modified dispersion relation for tensor modes when these operators are comparable to the Einstein Hilbert term. 

An important example of the former are Galileon theories~\cite{Nicolis:2008in}, which arise in flat space when demanding the field equations for a scalar field $\phi(x )$ to be invariant under the transformation $\phi\rightarrow\phi+bx$, with $b$  a constant.  This allows the introduction of  higher order derivative operators which, at the same time, are  second order in the equation of motion and finite.  Moreover the solution of the equations contains a self accelerating solution which can  replace  dark energy although these models are not compatible with observations of the distribution of large scale structure in the universe \cite{Appleby:2012ba}.   Allowing a curved background breaks the Galileon symmetry because there is not a covariant definition of shift symmetry. By requiring the theory to be second order, an approximate shift symmetry can be imposed and the Galileon theory can be \textit{covariantised}~\cite{Deffayet:2009wt, Deffayet:2009mn}. The resulting action also describes  the decoupling limit of massive gravity~\cite{deRham:2014zqa, deRham:2011by}. Also galileon gravity contain a consistent cosmological solutions which can be compared to observations~\cite{Chow:2009fm, Barreira:2012kk, Silva:2009km, Kobayashi:2010wa, DeFelice:2010nf, DeFelice:2010as, Nesseris:2010pc, DeFelice:2010pv, Brax:2014vla}.

 
 In this paper we will study  the speed of gravitons for Galileon theories on cosmological backgrounds. It turns out that the covariantisation procedure we outlined above introduces curvature interactions which also  modify  the graviton action at tree level. By studying this case we can also get a hint into possible generalisations such as  Horndeski theories or some limits of massive gravity~\cite{Horndeski:1974wa}. Moreover gravitational waves are simple  to model on a cosmological background and therefore more likely to be compared against observations. Although these theories have free parameters, the requirement of a physical cosmological background constrains them significantly. For example in~\cite{Brax:2015dma}, by assuming a Minkowski limit it was found that there are superluminal regimes for the graviton  which can be constrained by studying Cherenkov radiation from galactic sources. These  Galileon cosmologies have also been tested against CMB observations~\cite{Neveu:2013mfa, com},  interestingly in the CMB power spectrum of these models  the acoustic peaks are shifted compared to the standard cosmology case. 
 
The speed  of gravitons during inflation is not  necessarily one. Higher order curvature operators can easily modify the gravitational action~\cite{Baumann:2015xxa}. Also during Galileon inflation the curvature operators are enhanced and the speed is allowed to  change~\cite{Burrage:2010cu}. Nevertheless, as  was first pointed out in~\cite{Creminelli:2014wna},  deforming the metric  by a disformal transformation allows to set the speed of the gravitons equal to one. This is because disformal transformations change the slope of the lightcone, hence, by a suitable choice of parameters any speed can be removed as particles will move through  a modified light cone. This introduces a speed for the scalar field responsible of inflation, but due to the diffeomorphisms  invariance of the Einstein equations  this speed is not present in the curvature power spectrum and is a subleading effect on the bispectrum.  That being the case, then any  such correction to the speed of sound  is not observable. This result has been generalised to include Horndeski theories and therefore Galileon theories \cite{Tsujikawa:2014uza}. Also the more general case for the disformal invariance of the Einstein equations on a cosmological background was raised by~\cite{Domenech:2015hka}, where it was found that generally the disformal transformations leaves the equations unchanged.

In this work we revisit   the possibility of modifying the speed of light of tensor modes during inflation and at  later times. Despite the tight constraints from the observation of gravitational waves, it is  still valid to ask whether this speed is unobservable and can be removed by a disformal transformation as in inflation. Now the  situation during epochs where matter and radiation are dominating  can be  quite different  to inflation.   This is because during inflation all matter is washed out by de Sitter expansion and therefore  it is only necessary to consider the evolution of the inflaton. On the other hand during the matter and radiation domination epoch the  coupled  matter introduces characteristic time scales which may break time invariance.
More specifically we investigate the invariance of the Einstein equations and the Boltzmann equations under a disformal transformation.  It turns out that the change on the slope of the light cone changes the speed of propagation with respect to the untransformed  frame. Thus by a choice of parameters any speed appearing can be removed. Nevertheless, by assuming that the speed in either the tensor or scalar sector differs from one then, after the transformation  there will be an effective speed different from one for the other sector,  since the transformation affects the time and  scale factor.
We focus on the case where the graviton has an original speed of sound $c_T\neq 1$.  Removing it by deforming the light cone changes the speed of the scalar degree of freedom. 

In order to investigate whether the the CMB gets modified, we examine the evolution of photons coupled to gravity. Since the Boltzmann equations are  invariant,  the characteristic timescales of scattering processes that might occur do not change.  There is   a characteristic time scale, the Silk damping scale, characterizing the interactions of matter and photons   which is invariant under disformal transformation.  Therefore the comparison between the rates of these interactions and the graviton speed will not be invariant under a disformal transformation.  By solving the equations of motion the different speed of propagation shifts the peaks for the CMB anisotropy power spectrum, moreover because the Silk damping scale does not change compared to the untransformed case  the damping tail is different in comparison with Planck results. Hence, since any such modification leads to traceable effects, that modification cannot be applied to remove the speed of sound for the tensor modes.

It has been argued that  cosmological equations from Horndeski or beyond Horndeski theories are invariant under disformal transformation \cite{Domenech:2015hka}. As we argued before an approximate galileon symmetry can enhance curvature operator which allow the speed of  gravitons to vary. Hence by a disformal invariance this parameter might be removed from any observable.  Here we focus on Galileon theories because they capture the most important dynamics. We first show how curvature operators are enhanced during inflation leading to a modified speed for the gravitational waves. By using a disformal transformation this speed is removed from the tensor action, but it appears as a new parameter on the scalar action, we showed that this parameter dependence is removed from the power spectrum. By means of this symmetry the Galileon theory is further constrained. We also focus on Galileons during radiation domination, we showed  here that the speed for the gravitons cannot be removed because the transformation significantly changes the anisotropy power spectrum

The outline of this paper is as follows. We first explain what  disformal transformations  are and their  application to inflation. To do so we detail the EFT of inflation where the transformation can be better understood. We then calculate the transformation rules for the Einstein  and the Boltzmann equations. By analysing how matter couple to it we then show that the Silk damping scale is not modified, and thus when solving the coupled system its power spectrum will be damped at a different scale. Furthermore the new sound horizon change the peak structure and then small changes can be easily detected.  We illustrate this by solving the temperature anisotropy in the two fluids approximation~\cite{Seljak:1994yz}. While this is not very useful to get strong constraints it shows clear examples of how easy is to modify the power spectrum. 

Then, we  show how a non-unitary speed arises in Galileon theories on a cosmological background. We first detail Galileon inflation and calculate the speed for the tensor modes. By applying the disformal transformation we show explicitly  that this can be removed. We perform  a similar analysis for  Galileon theories during  radiation and matter domination. In this case the speed of the tensor modes is modified but observations of the CMB mean that  it is strongly constrained to be close to  one. Finally  we discuss our results and analyse its implications. 

\section{EFT of inflation}
\subsection{Scalar and tensor fluctuations}
To study modifications to the speed of light it is convenient to define an EFT because then it may be possible to organise the contributions to the two point function in terms of slow roll parameters. This can be done by using the framework first introduced in \cite{Cheung:2007st} where the realisation of inflation is parametrised as a a broken quasi de Sitter symmetry and the inflaton is  the resulting  pseudo Goldstone boson. Formally the action for  perturbations is given by, 
\bea
\label{1EFT:pert}
\Delta S=\int \mathrm{d}^4x\sqrt{-g}\left[\frac{M^4_2(t)}{2}(\delta g^{00})^2 +\frac{M^4_3(t)}{3!}(\delta g^{00})^3+\frac{M^4_4(t)}{4!}(\delta g^{00})^4+...\right.\ \nonumber\\
\left. -\frac{\bar{M}^3_1(t)}{2}\delta g^{00}\delta K-\frac{\bar{M}^3_2(t)}{2}(\delta K)^2-\frac{\bar{M}^3_3(t)}{2}(\delta K^\mu_{\ \nu})^2+...\right.\\
\left.-\frac{\hat{M}^2_1(t)}{2}\delta g^{00}R+...\right]\nonumber
\eea
Where $\delta g$ is the metric variation and $\delta K$ is the 3-curvature variation. We can see that contributions to the quadratic part of the gravitational section will be contained in the second order curvature operator.  Although one should expect them to be a perturbative series and then small compared to the first order operators, it can be the case that some higher order curvature operators are enhanced by a symmetry such as in DBI or Galileon theories.

The action for the perturbations will be given by first identifying  the Goldstone boson $\pi$ with the broken time invariance. Then this Goldstone boson will be the inflaton perturbation on a de Sitter background. In the limit where the mixing of $\pi$ with gravity is effectively decoupled the action is given by,
\bea
S&=&\int \mathrm{d}^4x\sqrt{-g}\left[\Mpl R-\Mpl\dot H\left(\dot\pi^2-\frac{(\partial\pi)^2}{a^2}\right)+2M^4_2\left(\dot\pi^2+\dot\pi^3-\dot\pi\frac{(\partial\pi)^2}{a^2}\right)-\frac{4}{3}2M^4_3\dot\pi^3...\right] \nonumber\\
\eea 
Note that that higher order operators modify the dispersion relation for the Goldstone boson. Indeed, the speed of sound can be written as
\bea
c_s^{-2}=1-\frac{2M_2^4}{\Mpl\dot H}
\eea
Moreover, $c_s$ parameterises the amount of equilateral non-Gaussianity, $f_{\rm{NL}}^{\rm{equil}}\propto c_s^{-2}$, as it relates how large $M_2^4$ is with respect to the first terms, hence it can be quantified how much the theory departs from a harmonic oscillator. In cases with more symmetry $c_s$ can be used to further constrain the theory.  For example in DBI inflation~\cite{Alishahiha:2004eh}, the next order parameter is found to be $M_3^4\sim\Mpl\vert\dot{H}\vert c_s^4$. Other interesting examples include effects coming from integrating out heavy degrees of freedom where it is possible that  the inflation interchanges  energy with other fields, then reducing its speed of propagation \cite{Achucarro:2010da}. 

The action for the gravitons is obtained by replacing the ansatz with  a traceless and divergenceless graviton $\de s^2=-\de t^2+a^2h_{ij}\de x^i\de x^j$ in (\ref{1EFT:pert}). We can decompose the Ricci scalar as, $R={}^{(3)}R +K_{ij}K^{ij}-K^2+..$. The first part yields the second order action,
\bea
S=\frac{\mpl^2}{8}\int\de^4x a^3\left[\dot\gamma_{ij}^2-\left(\partial \gamma_{kj}\right)^2\right].
\label{gravitons:action}
\eea
where higher order operators have the following contribution
\bea
(\delta K^\mu_\nu)^2-(\delta K)^2=\frac{1}{4}(\dot{\gamma}_{ij})^2
\eea
And thus only the combination $\bar{M}^3_3(t)-\bar{M}^3_2(t)$, can contribute to modifications of the speed of sound up to second order. The modifications for the speed of sound at first order on the slow roll parameters may arise from considering higher order curvature terms, such as the Weyl tensor squared $W^2$ or the parity violating Weyl tensor $\tilde{W}^2$~\cite{Baumann:2015xxa}. 

\subsection{Disformal transformation}
As we have seen modifications to the two point function for tensor and scalar modes can written terms of  an effective \textit{speed of sound} for each sector. Since this is the speed at which these degrees of freedom propagate through the lightcone, a redefinition of the null vectors could set  the speed  to one. This, also called  a disformal transformation~\cite{Bekenstein:1992pj},  is given by,
\bea
g_{\mu\nu}\mapsto \frac{1}{c}\left(g_{\mu\nu}+(1-c^2(t))n_\mu n_\nu\right)
\label{confdisf:transf}
\eea
Let us consider how the action for the EFT of inflationary transforms under (\ref{confdisf:transf}). We first calculate how it affects the background components,
\bea
S=\int\de^4 x\sqrt{-g}\left[\frac{\mpl^2}{2}R-\frac{\mpl^2}{c}\left(3H^2+\dot H\right)+\mpl^2\dot H g^{00}\right]
\label{EFTreplaced:action}
\eea
where  the  only change  is a re-scaling of Newton's constant, this is because the background is time dependent and thus it will be subject to a conformal rescaling which is equivalent to changing $\mpl$ in the Einstein equations. More interesting are the perturbations where gradients get a different rescaling. The second order action for the Goldstone boson $\pi$ becomes, 
\bea
S=\int\de^4 x\frac{a^3}{c^2}\mpl^2\dot H\left\{\dot{\pi}^2-\frac{c^2}{a^2}(\nabla\pi)^2\right\}
\label{EFTs:action}
\eea
Then it acquires a speed of propagation just due to the change of coordinates. We could have allowed a time dependent speed of sound, this would have included second order corrections but we are mostly interested in first order in slow roll parameters terms. 

\subsubsection*{Tensor modes}
We now consider the effect of the transformation (\ref{confdisf:transf}) on tensor perturbations. By means of this transformations the tensor modes changes as $h_{ij}\mapsto c^{-1}h_{ij}$. Therefore the second order action becomes, 
\bea
S=\frac{\mpl^2}{8}\int\de^4 x a^3\left[\frac{1}{c_T^2}\dot{\gamma}_{ij}^2 -\frac{(\partial_k\gamma_{ij})^2}{a^2}\right].
\label{ct3:action}
\eea
where again we see that gravitons  acquires a speed of sound from the rescaling of the parameters, similar to the action for the Goldstone boson $\pi$. We can also consider the corrections that give rise to a non-unity speed of sound. These rescale as
\bea
(\delta K^\mu_\nu)^2-(\delta K)^2=\frac{1}{4c}(\dot{\gamma}_{ij})^2
\eea
Then choosing that $\bar{M}^3_3(t)-\bar{M}^3_2(t)=(1-c^{-2})$ the previous action becomes,
\bea
S=\frac{\mpl^2}{8}\int\de^4 x a^3\left[\frac{1}{c^4}\dot{\gamma}_{ij}^2 -\frac{(\partial_k\gamma_{ij})^2}{a^2}\right].
\label{ct3modified:action}
\eea

\subsubsection*{Removing the speed of sound}
We have seen that the transformation (\ref{confdisf:transf}) sets a different speed of propagation for each mode. Therefore by choosing $c$ equal to $c_s$ or $c_t$ we can  remove  either of these  speeds and move it to an effective speed for the other sector.
\bea
\d s^2=-\de \tilde{t}^2+\tilde{a}(\tilde{t})\de \x^2,
\eea
where $\d \tilde{t}=c_T^{1/2}\d t$ and $\tilde{a}(\tilde{t})=c_T^{-1/2}a(t)$
For the transformed values the action becomes
\bea
S=\int\de^4 \tilde x\tilde a^3\mpl^2 \dot{\tilde{H}}\left\{c_T^2\dot{\tilde\pi}^2-\frac{1}{\tilde a^2}(\nabla\tilde\pi)^2\right\}
\eea
Now rewriting (\ref{ct3modified:action}) in terms of the rescaled variables we get that,
\bea
S_{\gamma\gamma}=\frac{\mpl^2}{8}\int\de\tilde t \de^3 x \tilde a^3\left[\dot{\tilde\gamma}_{ij}^2 -\frac{(\partial_k\tilde\gamma_{ij})^2}{a^2}\right]
\eea
Thus in the rescaled frame the graviton has speed of sound equal to one and the Goldstone mode propagates with a speed $c_s=1/c_T$. It remains to be seen however if these  modifications appear  in observables such as the power spectrum. This can be calculated by rescaling momenta by $k/c_s$, thus, 
\bea
\Delta_R\equiv \frac{k^3}{2\pi} P_\calR(k)=\frac{ c_T H^4}{4\mpl\dot H}
\eea
 In the transformed frame where $\tilde H=c_T H$ the power spectrum is equivalent to the case when $c_s=1$. Note that allowing  a $c_S\neq 1$ does not change the results as the next term in the EFT expansion gets the same rescaling than the tree level terms and then the second order action (\ref{EFTs:action}) just get a physical speed $c_S$, which cannot be removed. 

\section{Einstein equations}
We now want to see whether applying a disformal transformation can be used to remove any gravitons speed different from one throughout   cosmological evolution. One could rescale the time and the Hubble parameters from the Einstein equations but we prefer to keep track of all the speed  coefficients appearing to see when these can be safely removed and when they could not.  

Hence we first apply the disformal transformation to the perturbed Einstein equations in Newtonian gauge for both scalar and tensor sectors. It would be interesting to consider vector perturbations but since they decay fast during matter and radiation eras we will neglect them.  We start by considering the following transformed metric,
	\bea
ds^2=\frac{a^2(\tau)}{c_T}\left(-c_T^2(1+2\psi)\de\tau^2+(1-2\phi)\delta_{ij}\de x^i\de x^j\right)
\label{pert:metric}
\eea
which is the the perturbed FRW metric in Einstein frame under a disformal followed by a conformal transformation (\ref{confdisf:transf}). Assuming that the stress energy tensor is a perturbed perfect fluid, defined as,
\bea
T_{\mu\nu}=(\rho+P)u_\mu u_\nu+Pg_{\mu\nu}
\eea
where $u_\mu=\delta_{0\mu}/\sqrt{-g^{00}}$.  Since the perturbed stress energy tensor will have a spatial velocity we assume that this is of the form $\delta u^i=\frac{v^ic_T^{1/2}}{a}$, thus $\delta u^i\delta u_i=1$. Now the perturbed $4-$velocity is,
\bea
u_\mu=\frac{a}{c_T^{1/2}}(c_T(1+\psi),v_i).
\eea
Then we calculate the Einstein equations of motion. Its zero component are the Friedmann equations,
\bea
3\calH^2&=&8\pi Ga^2c_T\bar\rho\nonumber\\
-\calH^2-2\calH' &=& 8\pi Ga^2 c_T \bar{P}
\label{eqs:Friedmann}
\eea
which we see are dependent on the conformal factor as doing a conformal transformation is equivalent to reescaling Newton's constant. 
The Euler equations for the perfect fluid are,
\bea
\bar\rho'=-3\calH(\bar P+\bar \rho)
\eea
The perturbed equations, are given by

\bea
c_t^2\nabla^2\phi-3\calH(\calH\psi+\phi')=4\pi Ga^2c_T\delta\rho\nonumber\\
\calH\psi+\phi'=-4\pi Ga^2(\bar P+\bar \rho)v\nonumber\\
\phi''+\calH(\psi'+2\phi')+2\calH'\psi+\calH^2\psi-\frac{1}{3}c_T^2\nabla^2(\phi-\psi)=4\pi G a^2 c_T\delta P
\label{pertEinsten:eq}
\eea
where $v^i=\partial^i v$.

The perturbed Euler equation and continuity equations are,
\bea
\v'+3\calH(\frac{1}{3}-\frac{\bar P'}{\bar\rho'})\v&=&-\frac{c_T\nabla\delta P}{\bar\rho+\bar P}-c_T\nabla\psi\nonumber\\
\delta'+3\calH\left(\frac{\delta P}{\delta\rho}-\frac{\bar P}{\bar \rho}\right)\delta&=&-\left(1+\frac{\bar P}{\bar\rho}\right)(c_T\partial_i v^i-3\phi')
\label{PertEuler:eq}
\eea
Hence the disformal transformation also introduces a \textit{speed of sound} which can be different from one, for the perturbations in addition  rescaling the Newton constant. It can also be noticed that the second order in perturbations equations obtained by  combining (\ref{pertEinsten:eq}) and (\ref{PertEuler:eq}) will be wave equations with a a speed of propagation $c_T$.
It can also be remarked  that we could have obtained these results by doing a rescaling of time and the scale factor, but it is preferable to keep track of  where the $c_T$ factor appears. 

\subsection{Gravitational waves}
To calculate the power spectrum for gravitational waves we first apply the disformal transformation to the line element,
\bea
\de s^2=-a^2\d \tau^2+a^2\left(\delta_{ij}+2E_{ij}\right)\d x^i \d x^j,
\eea
where $E_{ij}$  parameterises the two tensor degrees of freedom, thus $E^i_i=E^i_{j,i}=0$, which becomes
\bea
\de s^2=-c_T a^2\d \tau^2+\frac{a^2}{c_T}\left(\delta_{ij}+2E_{ij}\right)\d x^i \d x^j.
\eea
Where similary to the case for the scalar perturbations (\ref{pert:metric}) the transformation factorised from the spatial and the temporal part of the line element.  Now we can calculate the perturbed Einstein tensor which is,
\bea
\delta G_{ij}=\frac{E''_{ij}}{c_T^2}-\nabla^2E_{ij}+\frac{2\calH E'_{ij}}{c_T^2}-\frac{2(2\calH'+\calH^2)E_{ij}}{c_T^2}
\label{pertGE:eq}
\eea
To calculate the first order perturbation to the stress energy tensor we notice that for the spatial part $u_i$ is first order in perturbation, therefore it will only contribute at second order in perturbations in the expansion of the stress energy tensor. Thus the only first order term is be given by,
\bea
\delta T_{ij}=\frac{2Pa^2E_{ij}}{c_T^2}.
\label{stressenergy:GW}
\eea
Summing both RHS of (\ref{pertGE:eq}) and (\ref{stressenergy:GW}) and using the background Einstein equations we get that
\bea
E''_{ij}-c_T^2\nabla^2E_{ij}+2\calH E'_{ij}=0,
\label{GWEinst:eq.}
\eea
which, as in inflation depends only on the Hubble parameter because we have not considered anisotropic matter. In order to do so we would have to modify (\ref{stressenergy:GW}) to include such contribution. This would have lead to a source term for the above equation.  Furthermore without matter the (\ref{GWEinst:eq.})   corresponds to a rescaling of  time and the scale factor, as for the scalar. 

\subsection{Modified speed of tensor modes}
Now that we know how the Einstein equations transform under a disformal transformation, the next question would be to applied this framework to the case when there is a modified $c_T$. We can imagine that by higher derivative corrections the gravitational waves equations get modified in such a way that there is an effective speed of sound,
\bea
E''_{ij}-c_T^2\nabla^2E_{ij}+2\calH E'_{ij}=0.
\label{GWEinstmod:eq.}
\eea
Which is the same equation as (\ref{GWEinst:eq.}), but in this case we have not applied any transformation to the metric. By applying an inverse disformal transformation
 we can cancel the $c_T$ out of the equations. Then equation (\ref{GWEinstmod:eq.}) can be written as,
\bea
\tilde E''_{ij}-\nabla^2\tilde E_{ij}+2\tilde \calH \tilde E'_{ij}=0.
\label{GWEinstset:eq.}
\eea 
Where we have expressed the functions in the new frame as tilded. 
This transformation will have an effect on the scalar modes, by inducing a tensor speed $c_S=c_T^{-1}$. For example the Einstein equations, (\ref{pertEinsten:eq}) become,
\bea
c_t^{-2}\nabla^2\tilde \phi-3\tilde \tilde \calH(\tilde \calH\tilde \psi+\tilde \phi')&=&4\pi G\tilde a^2c_T^{-1}\tilde \delta\rho\nonumber\\
\tilde \calH\tilde \psi+\tilde\phi'&=&-4\pi G\tilde a^2(\bar {\tilde P}+\bar{ \tilde \rho})\tilde v\nonumber\\
\tilde\phi''+\tilde\calH(\tilde \psi'+2\tilde\phi')+2\tilde\calH'\tilde\psi+\tilde\calH^2\tilde\psi-\frac{1}{3}c_T^{-2}\nabla^2(\tilde\phi-\tilde\psi)&=&4\pi G a^2 c_T^{-1}\delta\tilde P.
\label{pertEinstenset:eq}
\eea
To see whether there is an observable  effect we have to first solve these equations on a suitable background.
We will show that, unlike inflation observables will depend on $c_T$ because 
we need to consider how matter couples to the Einstein potential. 

\section{CMB}
To get an insight of the effect of $c_T$ we will solve Einstein equations to calculate the anisotropy power spectrum.  There it will be a   $c_T$ factor in the equations, because we   made a disformal transformation that rescaled the time and the scale factor, this will change  the particle horizon $\int \de t/a$ by a factor of $c_T$. 

Before recombination electron and baryons are tightly  coupled and they can be considered as a single fluid. This fluid will interact with photons through Thompson scattering, which is very efficient at large scales so it will be useful to consider a system of  two fluids, photons and baryons,  tightly  coupled, hence moving at the same speed. Since the Thomson cross section is finite there will be scales affected by it.  To take account of this    we also need transform the Boltzmann equation under a disformal transformation . 

We will solve the system numerically by using the two fluid approximation ~\cite{Seljak:1994yz} which solves the equations until the scales on which Thomson scattering is very efficient  and then interpolates to further epochs. This approach will be useful to get an insight on how important are the modifications introduced by a disformal transformation.

\subsection{Boltzmann equation}
To keep track of any modification to the propagation of  photons  at early times it will be convenient to calculate whether the Boltzman equation changes under a disformal transformation. This also includes a contribution coming from the the null geodesics which we need to take into account. Considering a perturbed perfect fluid with a energy perturbation  given by $E=\epsilon/a$, its  geodesic equation transforms under  (\ref{confdisf:transf}) as\footnote{See Appendix for details on the geodesic equations.},
\bea
\frac{\de\log \epsilon}{\de\eta}=-\frac{\de\psi}{\de\eta}+\psi'+\phi'
\eea
 where conformal time is now denoted by $\eta$. Note that any effect of the rescaling will be on the total derivative $\de/\de\eta=\partial_\eta+c_T e^i\partial_i$. Hence, there is no modification in the null geodesics which is in agreement to what one can expect as there it has not been a modification to the Einstein equations but a change of coordinates.
 
There is no change in the phase space due to the disformal transformations  because we have not changed the variables of the problem. Hence,  the Boltzmann equation  for the photon distribution $f$ is,
\bea
\frac{\de f}{\de \eta}=\frac{\partial f}{\partial \eta}+\frac{\partial f}{\partial x^i}\frac{\partial x^i}{\partial\eta}+\frac{\partial f}{\partial\log\eta}\frac{\partial\log\eta}{\partial\eta}=\frac{\de f_{\rm sct}}{\de \eta}\equiv\C[f,f_i]
\eea
where the RHS represents the scattering of photons with other species $f_i$, which we will detail later. Allowing only first order terms on $\epsilon$ we can reduce the Boltzmann equation to, 
\bea
\frac{\de f}{\de \eta}=\frac{\partial f}{\partial \eta}+c_Te^i\partial_i f+\frac{\partial f}{\partial\log\epsilon}\frac{\partial\log\epsilon}{\partial\eta}
\eea
Assuming that the background distribution is a black body, we have that at first order in the temperature perturbation $\Theta(\eta,\x)$,
\bea
f=\bar{f}(\epsilon)\left[1-\Theta(\eta,\x)\frac{\partial \bar f}{\partial\log\epsilon}\right].
\eea
Then up to first order the Boltzmann equation becomes,
\bea
\frac{\de \bar f}{\de\log\epsilon}\left(\frac{\partial\Theta}{\partial\eta}+c_Te^i\partial_i\Theta-\frac{\partial\log\epsilon}{\partial\eta}\right)=c_T\C[f,f_i]
\label{Boltzmann:eq}
\eea
Where the only modification arise in the spatial derivative for the temperature perturbations $\Theta(x)$. The dominant contribution to the scattering term will be the\textit{Thompson scattering}, which is the diffusion of  photons from the electron plasma. It is characterised by the Thomson cross section $\sigma_T=\frac{8\pi}{3}\left(\frac{q_e^2}{4\pi\epsilon_0 m_ec^2}\right)^2$.
The scattering term is not modified, because it relies in the Lorentz invariance of the distribution function, but the total contribution to the Boltzmann equation will be rescaled as $\de f_{\rm sct}/\de c_T\eta$. Hence the comoving mean free path is defined by $\Gamma=a\bar{n}_e\sigma_T$, where $\sigma_T $ is the \textit{Thomson} cross section, is invariant under the transformation. The  scattering rate in the expanding background can be written   as,
\bea
\C[f(\epsilon, \e)]=\frac{\de \bar{f}}{\de \ln {\epsilon}}\frac{\Gamma }{c_T}\left[\Theta-\e\cdot \v_\e+3\frac{1}{16\pi}\int\de \e_{\rm in}f(\epsilon, \e_{\rm in})[1+(\e_{\rm in}\cdot \e)^2]\right]
\eea
This last equation completes (\ref{Boltzmann:eq}) and the $c_T$ factor is absorbed in the space derivative. Hence the Boltzmann equation remains invariant under the disformal transformation. This will affect the propagation of the perturbation when their speed is other than one as Thomson scattering will introduce a different diffusion scale than in the untransformed case.

Now it is useful to decompose the temperature fluctuation  in terms of Legendre Polynomials $\Theta(\eta,\k,\e)=\sum_{l}(-i)^l\Theta_l(\eta,\k)P_l(\k\cdot \e)$. In which case the Boltzmann equation reduces to,
\bea
\frac{\de\Theta}{\de\eta}-\frac{\de\ln\epsilon}{\de\eta}=-\Gamma\left[\Theta-\Theta_0-i\mu v_e+\frac{1}{10}\Theta_2P_2(\mu)\right]
\eea
With $\mu$ the angle formed between the observation line and $v_e$. It can be shown that the first Legendre multipoles are related to the photon fluid variables as $\Theta_0=\frac{1}{4}\delta_\gamma$,  $\Theta_1=-v_\gamma$ and $\Theta_2=-\frac{5}{3}\sigma_\gamma$, thus we can relate last equation to the variables used in (\ref{pertEinsten:eq}) and (\ref{PertEuler:eq}). 

As we want to evaluate the perturbations at the surface of last scattering it is useful to define the  \textit{optical depth} along the light of sight between $\eta$ and $\eta_0$, 
\bea
\tau(\eta)\equiv\int^{\eta_0}_{\eta}\Gamma(\eta')\de\eta'
\eea
Which is invariant under disformal transformations because any factor coming from a rescaled time will be canceled with a corrected version of $\Gamma$
Using the line of sight parameterisation  and replacing the geodesic equation for the photon energy,  we then have that (\ref{Boltzmann:eq}) is rewritten as,
\bea
\frac{\de}{\de\eta}\left(e^{-\tau}(\Theta+\psi)\right)=S_{\mathrm{scal}}
\eea
where the scalar source term is
\bea
S_{\mathrm{scal}}=e^{-\tau}(\phi'+\psi')+g\psi+\frac{3}{16\pi}g\int\de\hat{\m}\Theta(\m)[1+(\e\cdot\hat{\m})]+g\e\cdot\v_b,
\label{ThomsonScattering}
\eea
where $g\equiv-\dot\tau e^{-\tau}$ and $\hat{\m}$ is a unit vector. Integrating along the line of sight and approximating last scattering as sharp, then the last equation reduces to,
\bea
\Theta(\eta_0,\x_0,\e)+\psi(\eta_0,\x_0)=\Theta_0+\psi+\e\cdot\v_b+\int^{\eta_0}_{\eta_*}\de\eta'(\psi'+\phi')
\eea
Where all terms are evaluated at the surface of last scattering $\eta_*$. Crucially, this last equation is invariant under disformal transformations. 

%

\subsection{Photon fluid}
We will consider a fluid composed of photons and baryons where electrons are tightly coupled to photons, hence  that two fluids will  move at the same speed. As baryons have no pressure we can define the two-fluid momentum as, 
\bea
\mathbf{q}=(\bar\rho_\gamma+\bar P_\gamma)\v_\gamma+\bar{\rho}_b\v_b\approx\frac{4}{3}(1+R)\bar{\rho}_\gamma\v_\gamma
\eea
where $R\equiv\bar{\rho}_b/(\bar\rho_\gamma+\bar P_\gamma)$ (using that $P/\rho=1/3$ for radiation). Considering that there are no anisotropic contributions to the stress energy tensors, the continuity equation (\ref{PertEuler:eq})  for baryons can be written as,
\bea
\v'_\gamma+\frac{\calH R}{1+R}\v_\gamma+\frac{c_T}{4(1+R)}\mathbf{\nabla}\delta_\gamma+c_T\nabla\psi&=&0\\
\delta_\gamma'+\frac{4}{3}c_T\nabla\cdot\v-4\phi'&=&0
\eea
Combining both equations we get
\bea
\delta_\gamma''+\frac{\calH R}{1+R}\delta_\gamma'-\frac{c_T^2}{3(1+R)}\mathbf{\nabla}^2\delta_\gamma=4\phi''+\frac{\calH R}{1+R}\phi'+\frac{4}{3}c_T^2\nabla^2\psi.
\label{photondistr:equation}
\eea
Which relates photon pressure to gravity effects. Note that there is a speed of sound $c_T$ induced on both photons and gravitational potential, which indicates that all scalar quantity are moving on a different lightcone. 
To solve this equation we need to solve first for the potential during radiation dominated epoch. Assuming a photon fluid with $a\propto \eta$, then the  equation of motion is
\bea
\phi''+\frac{5}{\eta}\phi'-\frac{c_T^2}{3}\nabla^2\phi=0
\eea
which has as solutions
\bea
\phi(\eta,\k)=A(\k)\frac{j_1(c_Tk\eta/\sqrt{3})}{c_Tk\eta/\sqrt{3}}+B(\k)\frac{n_1(c_Tk\eta/\sqrt{3})}{c_Tk\eta/\sqrt{3}}
\eea
Where $A(\k)$ and $B(\k)$ are dependent on the initial conditions, which we take to be  adiabatic initial conditions given by inflation.

\subsubsection*{Curvature perturbation}
In order to fix the initial conditions is better to express  the Newtonian potential $\phi$ in terms of the curvature perturbation $\calR$, which is constant out of the horizon. For  Newtonian gauge the relation is given by, 
\bea
\calR=-\phi-\frac{\calH v}{c_T}
\eea 
Where the $c_T$ factors comes  because the disformal transformation (\ref{confdisf:transf}) changes the spatial part of the gauge transformation.
Using the Einstein equations and the equation of state this can be expressed as
\bea
\calR=-\phi-\frac{2}{3(1+\omega)}\left(\psi+\frac{\phi'}{\calH}\right).
\label{curvaturePert:def}
\eea
Note that the dependence on $c_T$ has gone, as we are now considering a particular combination that should be invariant under coordinate transformations. Nevertheless, there is still an implicit dependence on $c_T$, which comes from the Newtonian potential.
To fix the initial conditions we will make some assumptions: First that $\calR(k)$ is adiabatic and scale invariant, also  we will not consider any contribution to the anisotropic stress energy tensor, hence $\phi=\psi$. Furthermore for modes outside of the horizon the last term is negligible, then we have,
\bea
\calR=-\frac{5+3\omega}{3+3\omega}\phi
\eea

\subsubsection*{Radiation domination}

For a photon fluid during radiation domination the above equation reduces  to $\calR=-\left(\frac{3}{2}\right)\phi$. Using this we can now  fix the initial conditions. For $k\eta\ll1$, since $\calR_k$ is constant outside the horizon, the solution for $\phi(k,\eta)$ is,
\bea
\phi(\eta,\k)=-2\calR_\k\frac{j_1(k\eta c_T/\sqrt 3)}{k\eta c_T/\sqrt 3},
\eea
which in the small scale limit ($k\eta\gg 1$) reduces to 
\bea
\phi(\eta,\k)=-6\calR_\k\frac{\cos(k\eta c_T/\sqrt{3})}{(k\eta c_T)^2}.
\eea
Note that there  will be a $c_T$ dependance on the evolution for the fields. 
To obtain the time evolution for the density we can use  (\ref{pertEinsten:eq}), which  for radiation domination  and using the background equations of motion becomes,
\bea
\delta_\gamma=-\frac{2}{3}\eta^2c_T^2k^2\phi-2\eta\dot\phi-2\phi.
\eea
Hence we have that for large scales $c_Tk\eta\ll1$, $\eta\dot\phi\ll\phi$,  the last equation simplifies to,
\bea
\delta_\gamma\approx-2\phi=\frac{1}{3}\calR_\k.
\eea
On large scales $c_Tk\eta\gg1$ only the first term dominates, thus,
\bea
\delta_\gamma(\eta,\k)\approx -\frac{2}{3}(kc_T\eta)^2\phi(\eta,\k)=-2\calR_\k\cos(k\eta c_T/\sqrt{3})
\eea

\subsection{Acoustic oscillations during matter domination}
Now we move to solve the Einstein equations for radiation during matter domination. First, during this epoch  the solution of equation (\ref{photondistr:equation}) is 
\bea
\delta_\gamma(\eta,\k)=C(\k)\cos(c_Tkr_S)+D(\k)\sin(c_Tkr_s)-4(1+R)\psi
\eea
where $r_s(\eta)=c_T\int^\eta_0 c_s(\eta')\de\eta'$ is the sound horizon, $c_S=\frac{1}{\sqrt{3(1+R)}}$, and $C(\k)$, $D(\k)$ are functions to be fixed by the initial conditions.  Note that the sound horizon is factorised by $c_T$ as now the horizon for a FRW is modified by the transformation. We could have also modified the speed of sound to include $c_T$, but for clarity we prefer to factor it out. 
From the definition of the curvature perturbation (\ref{curvaturePert:def}) during   matter domination we have that, 
\bea
\calR_\k=-\frac{5}{3}\phi
\eea
The Einstein equations imply that $\delta_\gamma-4\phi=\mathrm{const.}$ for all epochs. Using this    we can rewrite this as, $\delta_\gamma-4\phi=4\calR_\k$ for modes out of the horizon. Therefore to obtain $\delta_\gamma$ when  $kr_s=0$, we need to replace the value of $\phi$ during matter domination. Doing so results in 
\bea
\delta_\gamma&=&\frac{6}{5}\calR_\k,\nonumber\\
\dot\delta_\gamma&=&0,
\eea
which in turn implies that $D(\k)=0$ and, $C_k=-\frac{4(1+3R)}{5}\calR_\k$. Finally the density perturbation is,
\bea
\delta_\gamma(\eta,\k)=-\frac{4}{5}(1+3R)\calR_\k\cos(c_Tkr_s)+\frac{12(1+R)}{5}\calR_\k.
\eea
Using this we can calculate the Sachs Wolfe contribution to the anisotropy power spectrym. Using that, $\Theta_0=\delta_\gamma/4$,  we obtain, 
\bea
\Theta_0(\eta,k)+\psi(\eta,k)=-\frac{1}{5}\calR_\k\left[(1+3R)\cos(c_Tkr_s)-3R\right].
\label{SachsWolfeContr}
\eea
From the above equation we  have that  $\Theta_0(\eta,k)+\psi(\eta,k)$ oscillates between $-\frac{\calR_\k}{5}$ and  $(1+6R)\frac{\calR_\k}{5}$. So the amplitude of the peaks will not be changed due to the effect of $c_T$.

Now we can calculate the statistical anisotropies which  are given by,
\bea
\frac{l(l+1)C_l}{2\pi}\approx P_\calR(k)\left[\frac{\Theta_0(\eta,k)+\psi(\eta,k)}{\calR_k}\right]^2,
\eea
with $\Theta_0(\eta,k)+\psi(\eta,k)$ given by (\ref{SachsWolfeContr}). This means that the acoustic peaks will be given by $c_Tkr_s=n\pi$.  Therefore we should see a shift of them in the power spectrum due to the reescaling. 

\subsubsection*{Difussion damping}
At small scales Thomson scattering is not so efficient and the photons diffuse off the plasma. This can be calculated by relaxing the tight coupling assumption in (\ref{photondistr:equation}), in which case   $R$ is a now a  function of time. Considering that $R$ varies on cosmological time, thus $ R'=\calH R$ leads to a damped oscillator whose solution can be written as,~\cite{Hu:1994uz},
\bea
(\Theta+\psi)=(\Theta+\psi)e^{(k/k_D{\eta})^2}
\label{Diff:ansats}
\eea
where 
\bea
k_D^{-2}=\frac{1}{6}\int_0^\eta\de\eta\frac{1}{\Gamma}\frac{R^2+4(1+R)/5}{(1+R)^2}.
\label{Diff:scale}
\eea
To calculate how (\ref{Diff:ansats}) is modified under a disformal transformation  we can look at how  (\ref{photondistr:equation}) transformed, in which case (\ref{Diff:ansats}) is
\bea
(\Theta+\psi)=(\Theta+\psi)e^{(c_Tk/k_D{\eta})^2},
\eea
and (\ref{Diff:scale}) stays invariant. This is because Thomson scattering  relies on the same physics as in the untransformed case and all the changes due to the new speed are contained in the propagation of the perturbations. This different scaling will affect the damping of the anisotropy power spectrum as scales will be damped at $k/c_T$ respect to the untransformed case. 

\subsubsection*{Two fluids approximation}
\begin{figure}[!ht]
\begin{center}
\includegraphics[scale=0.5]{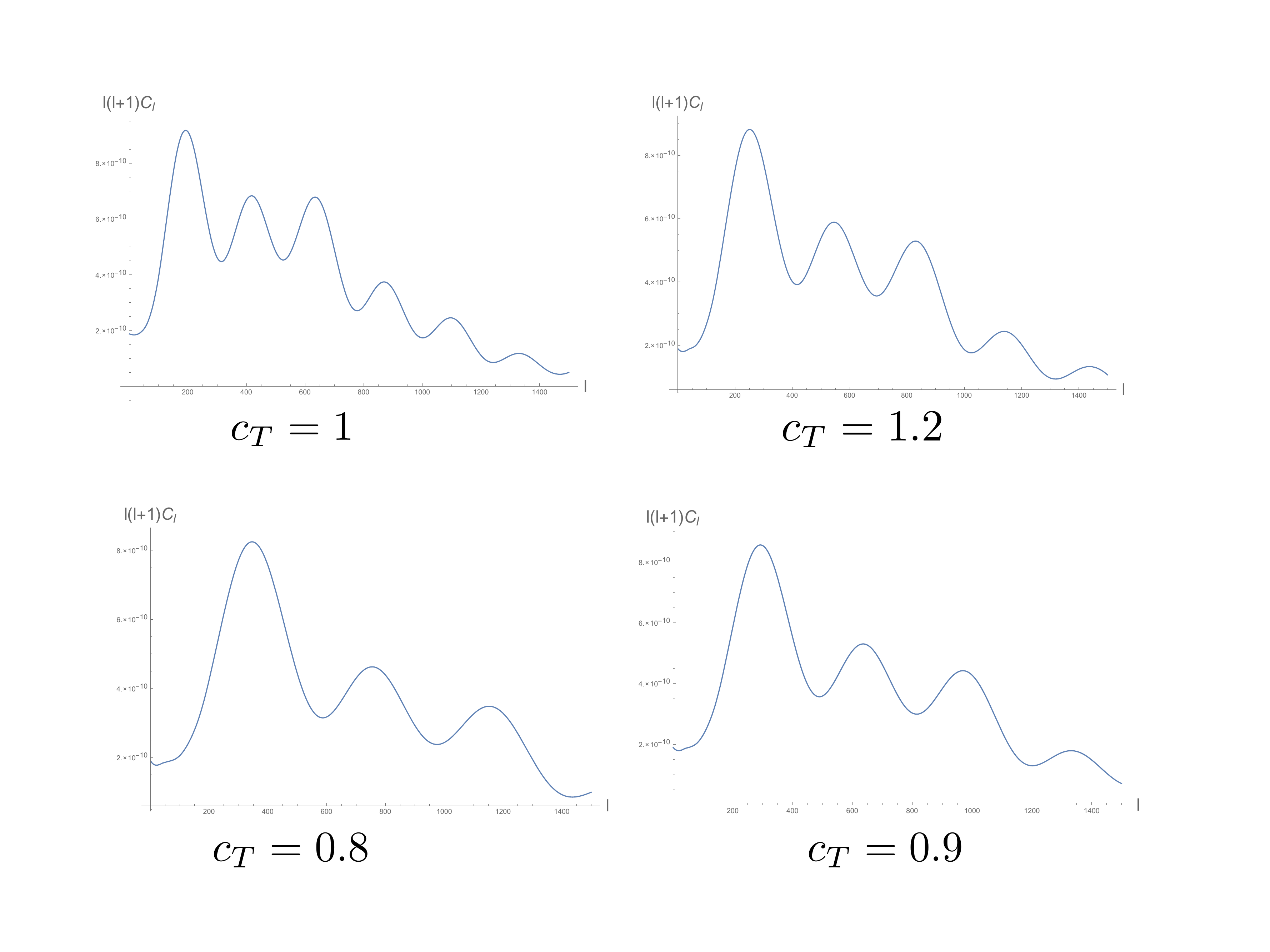}
\caption{$TT$ anisotropy power spectrum where the metric has undergo a disformal transformation (\ref{confdisf:transf}).}
\label{figure1}
\end{center}
\end{figure}
To get a better insight we will solve numerically the set of equations. We will use the two fluid approximation outlined  in~\cite{Seljak:1994yz}, where the equations for the tight coupling approximation are solved until the a cut off scale in which the approximation is still valid. Then this solution is extrapolated to smaller scales.  Finally the   the damping  factor \label{Diff:ansats} is  included which leads to an rough approximation from the correct power spectrum . This approximation it is no useful to estimate parameters but it can give us an idea of the effects that the disformal transformation on the power spectrum. 

In  Figure \ref{figure1} we have plotted  some cases where the speed varies around ten percent from one. Note how the position of peaks gets shifted as the dispersion relation  is modified. Furthermore, we see that the damping tail is modified. As we mentioned before, this can be understood  because all the modes gets a rescaling with the speed $c_Tk$. Therefore as modes reach the horizon at a different time they will be affected by the damping scale differently.


To summarise this section, we have shown that coupling matter to the Einstein equations does not allow us  to remove the speed of sound for the tensor modes by means of a disformal transformation. This is because  different scales are invariant under a disformal transformation, but  the observables are not invariant as they can trace information contained in the transformation.

\section{Galileons}
So far we have been  studying model-independent signatures that might arise when performing a disformal transformation, but it will be interesting to study what we can learn from applying this framework to theories where a $c_T\neq 1$ arise.  It is very hard to classify the large amount of theories that modify gravity and thus allow for a speed of tensor modes different from one, furthermore  some of these theories might have some other features that we have ignored, such as a non vanishing anisotropic stress energy tensor~\cite{Clifton:2011jh}. Nevertheless by requiring any theory to be able to give account of inflation and any later cosmological evolution highly restricts the class of available models.

We will now study gravitational waves on Galileon theories, which can embedded in inflation and also allows cosmological solutions~\cite{Burrage:2010cu,Chow:2009fm}. These are defined as the field equations invariant under the galileon transformation $\phi\rightarrow\phi+bx$ where $b$ is a constant.  They have a number of interesting properties but we will focus on the fact that the covariantised galileon theory allows large curvature operators to modify the speed of sound of the gravitational waves.   Since in this theory  the speed of the tensor modes is not necessarily one,  it will  demonstrate that the  disformal transformation is a helpful way to constrain and  better understand  theories of gravity. 

Furthermore     Galileon theories are related to the weakly coupled regime of massive gravity, and can be generalised to Horndeski theories. This allow us to understand analytically  the behaviour of higher order curvature corrections  in a large class of well motivated scenarios.

\subsection*{Galileon Action}
A curved background generally brakes the galileon symmetry producing higher order equation of motions.  Since we will focus  on cosmological backgrounds, it is necessary to consider an alternative to the original Galileon theories. 
Insisting on second order equations of motion, and accepting a breaking of the shift symmetry proportional to the background curvature yields the covariant formulation, whose action is~\cite{Deffayet:2009wt},
\bea
S\supseteq\int\de^4 x\sqrt{-g}\left[R+c_2\calL_2+c_3\calL_3+c_4\calL_4+c_5\calL_5\right]
\eea
where 
\bea
\calL_2&=&\frac{1}{2}(\nabla\phi)^2\\
\calL_3&=&\frac{1}{\Lambda^3}\square\phi(\nabla\phi)^2\\
\calL_4&=&\frac{(\nabla\phi)^2}{\Lambda^6}\left[(\square\phi)^2-\nabla_\mu\nabla_\nu\phi\nabla^\mu\nabla^\nu\phi-
\frac{R}{4}(\nabla\phi)^2\right]\\
\calL_5&=&\frac{(\nabla\phi)^2}{\Lambda^9}\left[(\square\phi)^3-
3\square\phi\nabla_\mu\nabla_\nu\phi\nabla^\mu\nabla^\nu\phi
+2\nabla_\mu\nabla_\nu\phi\nabla^\nu\nabla^\rho\phi\nabla_\rho\nabla^\mu\phi-G_{\mu\nu}\nabla^\mu\nabla\rho\nabla^\nu\nabla_\rho\phi\right]\nonumber\\
\eea
This action allows  a modified speed of sound for the tensor modes because there are second order corrections to the Ricci scalar in the quartic and quintic Galileon. 

We will study the propagations of gravitational waves in   two cases. First Galileon Inflation where the background has a broken  de Sitter symmetry which can be related to an inflationary epoch, and then we will study the Galileon on  a FRW background during matter and radiation eras. 
\subsection{Galileon inflation}
To allow inflation we consider a quasi de Sitter background representing an inflationary phase. Expanding in terms of the metric $\d s^2=-\d t^2+e^{Ht}d\x^2$ we have that,
\bea
S_0=\int\de^4xa^3\left[\frac{c_2}{2}\dot\phi^2+\frac{2c_3H}{\Lambda^3}\dot\phi^3+\frac{9c_4H^2}{2\Lambda^6}\dot\phi^4+\frac{6c_5H^3}{\Lambda^9}\dot\phi^5\right]
\label{DSGalileon:action}
\eea 
To study how non-linearities affect the dynamics it  is useful to define  the parameter $Z\equiv \frac{H\dot\phi}{\Lambda^3}$. Hence for $Z\ll 1$ there is a weakly coupled regime where the upshot is very similar to slow roll inflation. For $Z\gtrsim 1$ non-linearities are present and the inflationary dynamics is modified, thus a more careful analysis has to be done. This goes beyond the reach of this work.

To see how gravitons propagate we will focus on the the quartic Galileon. Expanding in terms of the metric, $\d s^2=-\d t^2+a(t)^2(\delta_{ij}+\gamma_{ij})\d x^i\d x^j$ we find that 
\bea
(\square\phi)^2-\nabla_\mu\nabla_\nu\phi\nabla^\mu\nabla^\nu\phi \subset -\dot{\phi}^2\frac{(\dot \gamma_{ij})^2}{4}
\eea
where the terms not included are proportional to the background equations of motion.  Then, substituting  back into  the action we find
\bea
S_g=\frac{1}{4}\int \d^4 xa^3(t)\sigma(t)\left[(\dot\gamma_{ij})^2 -c_T^2\frac{(\nabla \gamma_{ij})^2}{a^2}\right]
\eea
where,
\bea
2\sigma(t)&=&\Mpl-3c_4\dot\phi^4/\Lambda^2\nonumber\\
c_T^2&=&1+\frac{4c_4\dot\phi^2/\Lambda^2}{\Mpl-3c_4\dot\phi^2/\Lambda^2}
\eea
In the decoupling limit, where gravitational degrees of freedom can be ignored, we have that the action for small fluctuations can be obtained by using the Stuckelberg trick, 
\bea
S\supseteq\int\de^4 xa^3\left[\alpha\left(\dot\pi^2-\frac{c_s^2}{a^2}(\partial\pi)^2\right)+g_1\dot\pi^3+\frac{g_3}{a^2}\dot\pi(\partial\pi)^2+\frac{g_4}{a^4}(\partial\pi)^2\partial^2\pi\right]
\label{galfluctuations:action}
\eea
where $\alpha,\ c_s, \ g_1,\ g_2,\ g_3$ are functions of the background. As we are going to be interested in effects where there nonlinearities are small we can expand the above functions in terms of $Z$.  We will restrict our attention to $\alpha$ and $c_S$ which are
\bea
\alpha&=&\frac{\dot\phi^2}{2}\left(c_2+12c_3Z+54c_4Z^2+120c_5Z^3\right)\nonumber\\
c_s^2&=&1-\frac{4c_3}{c_2}Z+\left(\frac{48c_3^2}{c_2^2}-\frac{28c_4}{c_2}\right)Z^2
\eea
The wavefunctions can be solved in the case for small $Z$. Neglecting slow roll corrections, the power spectrum  for superhorizon scales is given by,
\bea
\Delta_\calR=\frac{1}{4\pi^3}\frac{H^4}{4\alpha c_s^3}
\label{Galileon:powerspectrum}
\eea
To summarise  during Galileon inflation there are terms which modify the dispersion relation in both the scalar and tensor sectors through its nonlinearities. As in slow roll inflation the graviton action is much simpler than its scalar counterpart. Although  only $c_4$ appears explicitly on the action, all the galileon dynamics will be present on the background functions. 

\subsubsection*{Conformal and disformal transformation}
 We will now study whether is possible to remove this tensor speed from the action through a disformal transformation,
\bea
g_{\mu\nu}\mapsto\frac{1}{c_T} \left(g_{\mu\nu}+(1-c_T^2(t))n_\mu n_\nu\right).
\label{confdisfgal:trans}
\eea
It is simpler if we first focus on the terms $(\nabla\varphi)^2$ and $\square\varphi$ to then derive how the Galileons actions transform. Under equation (\ref{confdisfgal:trans}) these become
\bea
g^{\mu\nu}\nabla_\mu\nabla_\nu\varphi&=&-\frac{\dot\varphi^2}{c_T}+c_Ta^{-2}(\partial_i\varphi)^2\\
\square\varphi=g^{\mu\nu}\nabla_\mu\nabla_\nu\varphi&=&-\frac{\ddot\varphi}{c_T}-\frac{3H}{c_T}\dot\varphi+a^{-2}c_T\partial^2\varphi.
\label{Galileonfields:transf}
\eea
Specialising to the case of a homogeneus and isotropic background the  Galileon action will then be reescaled as $\sqrt{-g}\calL_n\rightarrow \sqrt{-g}\calL_n/c_T^n$. The Galileon action for a de Sitter background then is written as,
\bea
S_0=\int\de^4xa^3\left[\frac{c_2}{2c_T^2}\dot\phi^2+\frac{2c_3H}{\Lambda^3c_T^3}\dot\phi^3+\frac{9c_4H^2}{2\Lambda^6c_T^4}\dot\phi^4+\frac{6c_5H^3}{\Lambda^9c_T^5}\dot\phi^5\right]
\label{ReescaledDSGalileon:action}
\eea 
To keep track of how the perturbations transform it  is useful to redefine $Z\equiv\frac{H\dot\phi}{\Lambda^3}$, as $\tilde Z=Z/c_T$. Now for the fluctuations 
the time derivatives add a $c_T^{-1/2}$ whereas spatial derivatives add $c_T^{1/2}$ to the corresponding terms in the Lagrangian.
Since we defined the perturbations by the map, $t\mapsto t+\pi(x)$, we  have that, $\phi\mapsto\phi+\dot\phi\pi$, hence the combination $\dot\phi\xi$ does not come with any factor of $c_T$. The coefficients for the Lagrangian (\ref{galfluctuations:action}) depends on the background as a function of  the parameter $\tilde Z=Z/c_T$. Then the second order  action for the perturbation is,
\bea
S\int\de^4xa^3\alpha\left[\frac{1}{c_T^2}\dot\pi^2-\frac{c_s^2}{a^2}(\partial\pi)^2\right]
\eea
where $\alpha$ and $c_s^2$ are now given by, 
\bea
\alpha&=&\frac{\dot\phi^2}{2}(c_2+12c_3\tilde Z+54c_4\tilde Z^2+120c_5\tilde Z^3)\\
c_s^2&=&1-\frac{4c_3}{c_2}\tilde Z+\left(\frac{48c_3^2}{c_2^2}-\frac{28c_4}{c_2}\right)\tilde Z^2+...
\eea
Following the same arguments the action for the gravitational waves is then given by, 
\bea
S_g=\frac{1}{4}\int \d^4 xa^3(t)\tilde\sigma(t)\left[\frac{(\dot\gamma_{ij})^2}{c_T^2} -\tilde c_T^2\frac{(\nabla \gamma_{ij})^2}{a^2}\right]
\eea
where,
\bea
2\tilde\sigma(t)&=&\Mpl-\frac{3c_4\dot\phi^4}{c_T^4\Lambda^2}\nonumber\\
\tilde c_T^2&=&1+\frac{4c_4\dot\phi^2/\Lambda^2}{c_T^4\Mpl-3c_4\dot\phi^2/\Lambda^2}
\label{tensorspeed:galileoninfl}
\eea
\subsubsection*{Removing $ c_T$}
Now, as in Section 1 we want to remove the physical $c_T$ by means of a conformal transformation. To do so we tune the parameter of (\ref{confdisfgal:trans}) to be the inverse of the tensor speed. Doing so, the tensor action becomes,
\bea
S_g=\frac{1}{4}\int \d^4 xa^3(t)\tilde\sigma(t)c_T^{2}\left[(\dot\gamma_{ij})^2 -\frac{(\nabla \gamma_{ij})^2}{a^2}\right]
\eea
whereas the scalar action (\ref{galfluctuations:action}) turns into,
\bea
S=\int\de^4xa^3\alpha c_T\left[\dot\pi^2-\frac{c_s^2}{c_T^2a^2}(\partial\pi)^2\right]
\eea
Now let us analyse this action. On the scalar part the background dependent functions gets corrected as  as $\alpha c_T$ and $c_s^2/c_T^2$, hence the power spectrum (\ref{Galileon:powerspectrum}) is now,
\bea
\tilde\Delta_\calR=\frac{1}{4\pi^3}\frac{\tilde H^4}{4\alpha  c_s^3c_T^{-2}}=\Delta_\calR,
\eea
because $\tilde H=c_s^{-1/2}H$. Therefore, as we indicated before, inflationary observables are invariant under this  transformation. Also this transformation imposes a constraint to the tensor speed as~\ref{tensorspeed:galileoninfl} becomes,
\bea
 c_T^2&=&1+\frac{4c_4\dot\phi^2/\Lambda^2}{c_T^4\Mpl-3c_4\dot\phi^2/\Lambda^2}
\eea
If we request  $c_T$ to be real, then $c_4\dot\phi^2/\Lambda^2>-10^{-2}\Mpl$. Since the cut off scale is $\Lambda\sim \phi$, the former bound can be rewritten as $c_4 H^2>-0.01\Mpl$ . This strongly constraints negative values for $c_4$. This is interesting as having a positive $c_4$ leads to superluminal propagation.
\subsection{Galileons on a cosmological background}
Now we would like to investigate how Galileon symmetry modifies the speed of tensor modes during the  matter and radiation epochs. We will consider the quartic Galileon as we are interested in how the Galileon symmetry can  enhance operators which modify the dispersion relation for the tensor modes. 

During matter and radiation era the Galileon component will modify the background through the Einstein equation $3\calH^2=8\pi Ga^2\sum_i\bar{\rho}^i$, where $i$ counts all the matter species which in this case includes the Gaileon density as well as  matter and radiation. Hence we need to consider a suitable evolution for the Galileon during this epoch. In the case of a quartic Galileon this is given by imposing that the galileon density dominates at late times.

\subsubsection*{Tensor modes}
For   a  FRW background $\d s^2=-\d t^2+a^2\d\x^2$ the  equations for tensor fluctuations will be very similar to those for  inflation because any matter component will be coupled only to scalars. Therefore we have that for metric perturbations given by the line element $\de s^2=-a^2\de \tau^2+2a^2E_{ij}\de x^i\de x^j$,   the Einstein equations simplifies to,  
\bea
E''_{ij}-c^2_T\nabla^2E_{ij}+2\calH E'_{ij}
\eea
where 
\bea
c_T^2&=&1+\frac{4c_4\dot\phi^2/\Lambda^2}{\Mpl-3c_4\dot\phi^2/\Lambda^2}.
\label{gravwavesspeed:galileon}
\eea
Hence, as a result of the curvature operators allowed in the Galileon theory   we have a modified speed of sound. When confronted with experiments we could have expect to get strong constraints for $c_4$. We can instead, try to keep $c_T$ large by removing it from this action by a disformal transformation. 
\subsubsection*{Scalar modes}

We need to obtain the perturbation equations during matter and radiation epochs,  assuming that Galileons are only coupled to matter. Moreover we will be interested in the case of a dominating quartic Galileon $c_4$, hence there it will not be anisotropic stress energy tensor and $\phi=\psi$. We can then modify what we did in Section 3 for scalar perturbations because the Galileons will act a source term to (\ref{PertEuler:eq}), while their equation given by varying~(\ref{DSGalileon:action}). At late time this yields a self accelerating solution, but we are more interested in the case before decoupling. Then we need to see whether there is a modification to the perturbation growth equation at early times. There are many ways of parameterising deviations from GR. Let us follow the approach of~\cite{Burrage:2015lla}, where the Galileon  contributions are parameterised as fifth-forces. It follows that the continuity equations~(\ref{PertEuler:eq}) become,
\bea
\nabla\cdot\v_m'+3\calH\nabla\cdot\v_m+\nabla\psi&=&F_\phi\\
\delta'+(\nabla\cdot\v_m-3\phi')&=&j_\phi
\eea
Where $F_\phi$ and $j_\phi$ are functions of the Galileon background only. In the case where $F_\phi$ is proportional to $\nabla\cdot\v_m$, as it is for the quartic Galileon we can redefine the Hubble parameter. Hence the equation becomes,
\bea
\nabla\cdot\v_m'+3\tilde\calH\nabla\cdot\v_m+\nabla\psi=0
\eea
With this in mind we can now build the second order equation. Since there are effects on the background which  are similar for both scalar and tensor modes it is useful to write the equations in such a way that we can identify what corresponds to the disformal transformation. We wrote a two fluid equation (\ref{photondistr:equation}), which demonstrated that  the speed for the photon density was slowed down because of baryons. The equation now becomes
\bea
\delta_\gamma''+\frac{\tilde\calH R}{1+R}\delta_\gamma'-\frac{1}{3(1+R)}\mathbf{\nabla}^2\delta_\gamma=4\phi''+\frac{\tilde\calH R}{1+R}\phi'+\frac{4}{3}\nabla^2\psi
\label{Galileonphotondistr:equation}
\eea

It follows from this, that the main contribution from a Galileon background will be a different Hubble parameter. Whereas this effect will certainly affect the solution of   equation~(\ref{Galileonphotondistr:equation}), it does not change the   speed of the scalar  and density perturbations $\phi$ and $\delta_\gamma$. Moreover a change on the Hubble parameter can be removed by redefining  Newton's constant through Equation (\ref{eqs:Friedmann}). This will set a new scale for the whole set of Einstein equations but the scalar perturbation will still have speed one on the new frame.  

Therefore we can apply the conformal transformation and  rely on the results we have obtained for inflation for the speed of the tensor modes. Since the transformation will rescale a subdominant contribution to the Hubble scale we can neglect this effect. The transformed equation (\ref{Galileonphotondistr:equation}) is,
\bea
\delta_\gamma''+\frac{\tilde\calH R}{1+R}\delta_\gamma'-\frac{c_T^2}{3(1+R)}\mathbf{\nabla}^2\delta_\gamma=4\phi''+\frac{\tilde\calH R}{1+R}\phi'+\frac{4}{3}c_T^2\nabla^2\psi
\label{Galileonphotondistr:equation2}
\eea
With this we can use what we detailed in Section 4. Then, the disformal transformation will change the temperature power spectrum in a significant way and the physical speed of sound cannot be removed from the tensor modes. Although this result assumes that the galileon evolution does not affect significantly the Einstein equations this is a valid assumption to make, as the matter and radiation  considered is well within the Vainshtein radius, and thus significant effects are screened. On the other hand we have allowed  $c_4$ to be large compared to the other terms $c_3$ and $c_2$. Although this does not holds in general, it allows us  to study the effect of the disformal transformation more transparently as we do not have to take into account the complicated equation that these terms induce. 

In any case it will be interested to study numerically how these terms are constrained due to requirement of $c_T^2\approx 1$. For example (\ref{gravwavesspeed:galileon}) reduces to 
$\vert c_4\dot\phi^2/\Lambda^2\vert\lesssim 10^{-7}\Mpl$.


\section{Conclusions}
We have studied how gravitational waves can vary their speed during different cosmological epochs. This is an interesting question as modifications to the speed of sound for tensor modes can arise in a large class of models. As there are strong constraints on the speed of the tensor modes we have focus on the possibility to set it to 1 by means of a disformal transformation. This has been shown to be possible in inflation but we have shown that a later times this  is more subtle. 

By studying the equations that give rise to the CMB we have shown that the anisotropy power spectrum contains trace of a disformal transformations, more precisely the scalar potential  acts as if it were propagating at a speed other than one. This effect is present in the baryoacoustics oscillations and also in the diffusion damping.  On the first it shifts the acoustic peaks as the modes propagates at a different speed. The relevant scales for the damping are invariant under a transformation, hence the scales are affected differently as modes became smaller or larger differently compared to the horizon size. 

This result is interesting and it can be used to constraint  the role of higher curvature operators. We have shown that they can arise in galileon theories because there higher order terms, containing corrections to the Ricci scalar, which can be very large. By using this we have found bounds on the parameter space which can shed some light on the realisation of a large class of theories.

There are many other questions to continue this work. We have study analytically how the anisotropy power spectrum varies when a disformal transformation is made. but very strong bounds to $c_T$ can be obtained from cosmological data . Furthermore by using this bounds the parameter space of galileon theories can be constrained. Also there are other theories which present similar characteristic and it will be interesting to study them using the methods we have discussed in this work.
\section*{Acknowledgements} CB is supported by a Royal Society University Research Fellowship. SC is supported by  Conicyt through its programme Becas Chile and the Cambridge overseas Trust. SC and ACD  acknowledges partial support from STFC under grants ST/L000385/1 and ST/L000636/1. SC would like to thank Gonzalo Palma for helpful discussions.


\appendix
\section{EFT of inflation}
A good choice to study perturbations on a cosmological background is by foliating the spacetime into hypersurfaces of constant time. This naturally introduces the ADM decomposition where we have that, 
\bea
\de s^2=-N^2\de t^2+h_{ij}(N^i\de t+\de x^i)(N^j\de t+\de x^j)
\eea
with functions $N$ and $N^i$ called lapse and shift, parameterising time evolution and its projection on the hypersurface. They are constraint which value is  fixed by the Einstein equations, when imposing a metric, eg.  a FRW universe. For this case, furthermore,  at zero order they may  only be time dependent because they  are functions of  an isotropic and homogeneous solution. Since we will perturbe the metric slightly, we can build order parameters, based on background quantities, which allow to have control on how  accurate is our description.

We can think of inflation as a quasi de Sitter spacetime  which has broken time diffemorphism and because it ends. Then, an effective field theory (EFT) for deviations from this broken de Sitter spacetime can be written in term of the remaining symmetries of this spacetime.  In the foliation that we described, functions of the induced metric $h_{\mu\nu}$ on the hypersurface $\Sigma_t$ will be allowed, such as the the the extrinsic curvature tensor $K_{\mu\nu}=h_\mu^\rho\nabla_\rho n_\nu$, and the Riemman curvature $\hat{R}_{\alpha\beta\gamma\delta}$ on the induced metric $\Sigma_t$. Also products of four dimensional covariant tensors with free upper 0 indices, but with all spatial indices contracted are allowed operators ($g^{00}$, $R^{00}$)

The most general action constructed with these elements is 
\bea
S=\int \mathrm{d}^4x\sqrt{-g}\calL\left[R_{\mu\nu\rho\sigma},g^{00},K_{\mu\nu},\hat{R}_{\mu\nu},\nabla_\mu;t\right]
\label{ansatz:action}
\eea
\subsection{Scalar action}
Since we are describing perturbations around a given background, the time dependent part of this action will be a tadpole term equivalent to the Friedmann equations.
\bea
S=\int \mathrm{d}^4x\sqrt{-g}\left[\frac{\mpl^2}{2}R+\Mpl(3H^2+\dot{H})+\Mpl\dot{H}g^{00}\right]
\eea
 The rest can be described in term of the fluctuations. After fixing the gauge invariance, the full action will be described in term of two independent functions. In the Newtonian gauge, where the hypersurface of constant time $\Sigma_T$ are orthogonal to the wordiness of observers at rest, the perturbations are described as functions of $\delta g^{00}=g^{00}+1$ and $\delta K_{\mu\nu}=K_{\mu\nu}-Hh_{\mu\nu}$. The action is formally

\bea
\label{EFT:pert}
\Delta S=\int \mathrm{d}^4x\sqrt{-g}\left[\frac{M^4_2(t)}{2}(\delta g^{00})^2 +\frac{M^4_3(t)}{3!}(\delta g^{00})^3+\frac{M^4_4(t)}{4!}(\delta g^{00})^4+...\right.\ \nonumber\\
\left. -\frac{\bar{M}^3_1(t)}{2}\delta g^{00}\delta K-\frac{\bar{M}^3_2(t)}{2}(\delta K)^2-\frac{\bar{M}^3_3(t)}{2}(\delta K^\mu_{\ \nu})^2+...\right.\\
\left.-\frac{\hat{M}^2_1(t)}{2}\delta g^{00}R+...\right]\nonumber
\eea
where the time dependent functions are function of the background, and vary slowly on time. Furthermore for slow roll inflation, they will be functions of the slow roll parameters. By reintroducing time invariance by means of the transformation
\bea
t&\mapsto& t'=t+\pi(t,x^i)\\
x^i&\mapsto& x_i'=x_i
\eea
where $\pi(t,x^i)$ is the Goldstone boson, the action for the scalar part can be extracted from the EFT by identfying $\pi$, with the inflaton. Then, the action (\ref{EFT:pert}), becomes,
\bea
S&=&\int \mathrm{d}^4x\sqrt{-g}\left[\frac{\mpl^2}{2}R+\Mpl(3H^2(t+\pi)+\dot{H}(t+\pi))+\right.\\
&&\left.\hspace{2.2cm}+\Mpl\dot{H}(t+\pi)(g^{00}+2\partial_\mu\pi g^{0\mu}+\partial_\mu\pi\partial_\nu\pi g^{\mu\nu})\right.\\
&&\left.\hspace{2.2cm} +\sum_n \frac{M_n^4(t+\pi)}{n!}(1+g^{00}+2\partial_\mu\pi g^{0\mu}+\partial_\mu\pi\partial_\nu\pi g^{\mu\nu})^n\right]
\eea
where as we have anticipated the first function depends on the slow roll parameter $\epsilon=- \frac{\dot H}{H^2}$ and thus vary slowly on time. 
To further simplify it is useful to take the limit  where Goldstone modes decouple from gravity. This happens where,
\bea
\mpl\rightarrow \infty,\hspace{1cm}\dot{H}\rightarrow 0\hspace{1cm}\mpl^2\dot{H}=const,
\eea
Then we have to evaluate if mixed terms are small compared to the kinetic energy $\mpl^2\dot{H}(\partial_\mu\pi)^2$. For example at background level the leading mixing with gravity comes from the term $\Mpl\dot H\delta g^{00}$
Then the action becomes
\bea
S&=&\int \mathrm{d}^4x\sqrt{-g}\left[\frac{\mpl^2}{2}R+\Mpl(3H^2(t+\pi)+\dot{H}(t+\pi))+\right.\nonumber\\
&&\left.\hspace{2.2cm}+\Mpl\dot{H}(t+\pi)(-1-2\dot\pi^2+(\partial_\mu\pi)^2)\right.\nonumber\\
&&\left.\hspace{2.2cm} +\sum_n \frac{M_n^4(t+\pi)}{n!}(-2\dot\pi^2+(\partial_\mu\pi)^2)^n\right]
\label{decoupled:action}
\eea
Up to third order we have  that,
\bea
S&=&\int \mathrm{d}^4x\sqrt{-g}\left[\Mpl R-\Mpl\dot H\left(\dot\pi^2-\frac{(\partial\pi)^2}{a^2}\right)+2M^4_2\left(\dot\pi^2+\dot\pi^3-\dot\pi\frac{(\partial\pi)^2}{a^2}\right)-\frac{4}{3}2M^4_3\dot\pi^3...\right] \nonumber\\
\eea 
Note that that higher order operators modify the dispersion relation for the Goldstone boson. Indeed, the speed of sound can be written as
\bea
c_s^{-2}=1-\frac{2M_2^4}{\Mpl\dot H}
\eea
Where this can vary from one by a different set of effects as we described in section 2.
\subsection{Tensor fluctuations}
We define the scalar perturbations as $\zeta$ and tensor perturbations as $\gamma_{ij}$ as
\bea
h_{ij}=a^2e^{2\zeta}(e^{\gamma})_{ij} \hspace{1cm}\mathrm{with}\hspace{1cm}\gamma_{ii}=\partial_i\gamma_{ij}=0
\eea
We will be interested in corrections to the two point function. Then the relevant contribution  to the  second order equation of motion for the tensor modes comes from  $M^4_2(t)$, $\bar{M}^3_1(t)$, $\bar{M}^3_2(t)$ and $\bar{M}^3_3(t)$. Expanding in the usual ADM decomposition, 
\bea
\de s^2=-N^2\de t^2+h_{ij}(N^i\de t+\de x^i)(N^j\de t+\de x^j)
\eea
where the extrinsic curvature is given by $K_{ij}=\frac{1}{2N}(\dot{h}_{ij}-\nabla_i N_j-\nabla_j N_i)$. Now expanding we will have that when looking to tensor perturbations at first order in curvature perturbations $N^i=0$ and $N=1$. In ADM the Ricci scalar is decomposed as,
\bea
R={}^{(3)}R +K_{ij}K^{ij}-K^2+...
\eea
where the dots corresponds to boundary terms. We then have that up to second order in $h_{ij}$
\bea
{}^{(3)}R&=&-\frac{1}{4a^2}\left(\partial \gamma_{kj}\right)^2\\
K_{ij}K^{ij}-K^2&=&-6H^2+\frac{1}{4}\left(\dot\gamma_{ij}\right)^2
\eea
And then the background action for the gravitons is
\bea
S=\frac{\mpl^2}{8}\int\de^4x a^3\left[\dot\gamma_{ij}^2-\left(\partial \gamma_{kj}\right)^2\right].
\label{gravitons:action}
\eea
Furthermore we have that
\bea
(\delta K^\mu_\nu)^2-(\delta K)^2=\frac{1}{4}(\dot{\gamma}_{ij})^2
\eea
Hence the combination of terms given by $\bar{M}^3_3(t)-\bar{M}^3_2(t)$, can modify the speed of sound up to second order in perturbations.
\section{Photons geodesics}
\label{Photons distribution}
We have to find the geodesic equations for photons. First 
we parametrise the   photons momentum as,
\bea
p^{\mu}=E(E_0^{\mu})+p^i(E_i^\mu),
\eea
We also define
\bea
E=\epsilon/a, \hspace{2cm} p^i=\frac{\epsilon}{a}e^i \hspace{0.5cm}\mathrm{with}\hspace{0.35cm} \delta_{ij}e^i e^j=1
\eea
given that 
\bea
g_{\mu\nu}E_0^\mu E_0^\nu=-1\hspace{1cm}\mathrm{and}\hspace{1cm}g_{\mu\nu}E_i^\mu E_j^\nu=\delta_{ij},
\eea
we have thus,
\bea
E_0^0&=&\frac{1}{a\sqrt{c_T}}(1-\psi)\\
E_i^i&=&\frac{\sqrt{c_T}}{a}(1+\phi)
\eea
then,
\bea
p^\mu=\frac{\epsilon}{a^2}\left[\frac{1-\psi}{\sqrt{c_T}},\sqrt{c_T}(1+\phi)e^i\right],
\label{phmom:def}
\eea
 which satisfies the constraint $p^2=0$. The geodesic equations are,
\bea
\frac{\de x^\mu}{\de \lambda}=p^\mu, \hspace{2cm} \frac{\de p^\mu}{\de \lambda}+\Gamma^{\mu}_{\alpha\beta}p^{\alpha}p^{\beta}=0
\eea
Replacing using (\ref{phmom:def}) it follows,
\bea
\frac{\de x^0}{\de \lambda}&=&\frac{\epsilon}{a^2}\frac{1-\psi}{\sqrt{c_T}}\\
\frac{\de x^i}{\de \eta}&=&\frac{p^i}{p^0}=c_T(1+\phi+\psi)e^i
\eea
The geodesic equations for the 0 component is ,
\bea
\frac{\de\log \epsilon}{\de\eta}=\frac{\de\psi}{\de\eta}-\psi'+\phi'-2c_Te^i\partial_i\psi 
\label{photongeo:eq}
\eea
Noticing that de derivative of $\psi$ is along the photon trajectory and thus it is written as,
\bea
\frac{\de\psi}{\de\eta}&=&\frac{\partial\psi}{\partial\eta}+\frac{\partial\psi}{\partial x^i}\frac{\partial x^i}{\partial \lambda}\frac{\partial\lambda}{\partial\eta}\nonumber\\
&=&\psi'+c_Te^i\partial_i\psi
\eea
The last equation becomes,
\bea
\frac{\de\log \epsilon}{\de\eta}=-\frac{\de\psi}{\de\eta}+\psi'+\phi'
\eea

where let us notice that the derivatives of the temperature perturbation are along the photon trajectory and thus the last equation can be written as, 

\bea
\frac{\partial \bar f}{\partial\log\epsilon}\frac{\de}{\de\eta}\left(\Theta-\log\epsilon\right)=\frac{\de f}{\de \eta}_{sct}
\eea
\bibliographystyle{jhep}
\bibliography{disformalCMB} 

\end{document}